\documentstyle[12pt]{article}

\makeatletter

\newfont{\Bbb}{msbm10 scaled 1\@ptsize00}

\newcommand{\ZZ}{\mbox{\Bbb Z}}
\newcommand{\RR}{\mbox{\Bbb R}}

\newif\if@fewtab\@fewtabtrue

\def\draftdate{\number\day.\number\month.\number\year\ \ \ \hourmin }
{\count255=\time\divide\count255 by 60
\xdef\hourmin{\number\count255}
\multiply\count255 by-60\advance\count255 by\time
\xdef\hourmin{\hourmin:\ifnum\count255<10 0\fi\the\count255}}
\def\ps@draft{\let\@mkboth\@gobbletwo
    \def\@oddhead{}
    \def\@oddfoot
       {\hbox to 7 cm{$\scriptstyle\bf Draft\ version:\ \draftdate$
       \hfil}\hskip -7cm\hfil\rm\thepage \hfil}
    \def\@evenhead{}\let\@evenfoot\@oddfoot}

\def\label#1{\ifnum\draftcontrol=1
 \global\def\draftnote{\scriptsize\tt #1}\fi
 \@bsphack\if@filesw {\let\thepage\relax
   \def\protect{\noexpand\noexpand\noexpand}%
\xdef\@gtempa{\write\@auxout{\string
      \newlabel{#1}{{\@currentlabel}{\thepage}}}}}\@gtempa
   \if@nobreak \ifvmode\nobreak\fi\fi\fi
  \@esphack}

\def\@eqnnum{\hbox to 3cm{\phantom{\rm(\theequation)} \draftnote
                         \hfil}\hskip -3cm {\rm(\theequation)}}

\def\eqnarray{\def\draftnote{{}}\global\@fewtabtrue
\stepcounter{equation}\let\@currentlabel=\theequation
\global\@eqnswtrue
\global\@eqcnt\z@\tabskip\@centering\let\\=\@eqncr
$$\halign to \displaywidth\bgroup\@eqnsel\hskip\@centering\@eqcnt\z@
  $\displaystyle\tabskip\z@{##}$&\global\@eqcnt\@ne
  \hskip 1\arraycolsep \hfil${##}$\hfil
  &\global\@eqcnt\tw@ \hskip 1\arraycolsep
$\displaystyle\tabskip\z@{##}$
\hfil  \tabskip\@centering&\global\@eqcnt\thr@@\llap{##}\tabskip\z@
\cr}

\def\endeqnarray{\@@eqncr\egroup
      \global\advance\c@equation\m@ne$$\global\@ignoretrue}

\def\@@eqncr{\let\@tempa\relax
    \ifcase\@eqcnt \def\@tempa{& & &}\or \def\@tempa{& &}
      \or \def\@tempa{&}
      \or\def\@tempa{}
\fi\@tempa
\if@eqnsw
\if@fewtab\@eqnnum\fi
\stepcounter{equation}\fi\global
\@eqnswtrue\global\@eqcnt\z@\global\@fewtabtrue\cr}

\@addtoreset{equation}{section}

\def\cases#1{\left\{\,\vcenter{\normalbaselines\m@th
    \ialign{$\displaystyle{##}\hfil$&\quad##\hfil\crcr#1\crcr}}\right.}

\def\ct#1{\ifnum\draftcontrol=1{\tt [#1]}\else{\cite{#1}}\fi}
\def\ctz#1#2{\ifnum\draftcontrol=1{\tt [#1,#2]}\else{\cite[#1]{#2}}\fi}

\def\draftcite#1{\ifnum\draftcontrol=1#1\else{}\fi}

\def\@lbibitem[#1]#2{\item{}\hskip -3cm \hbox to 2cm
{\hfil$\scriptstyle\draftcite{#2}$}\hskip
1cm[\@biblabel{#1}]\if@filesw
     {\def\protect##1{\string ##1\space}\immediate
      \write\@auxout{\string\bibcite{#2}{#1}}}\fi\ignorespaces}

\def\@bibitem#1{\item\hskip -3cm \hbox to 2cm
{\hfil \scriptsize\tt\draftcite{#1}}\hskip 1cm
\if@filesw \immediate\write\@auxout
       {\string\bibcite{#1}{\the\value{\@listctr}}}\fi\ignorespaces}

\makeatother

\def\lb#1{\label{#1}}
\def\lab#1{\ifnum\draftcontrol=1{{\tt [#1]} \lb{#1}}\else{\lb{#1}}\fi}
\def\Eq#1{(\ref{#1})}
\def\theequation{{\thesection.\arabic{equation}}}
\def\[{\begin{eqnarray}}

\def\non{\nonumber \\ }
\def\]{\end{eqnarray}}

\def\een{\end{enumerate}}
\def\ben{\begin{enumerate}}

\renewcommand{\a}{\alpha}
\renewcommand{\b}{\beta}
\renewcommand{\c}{\gamma}

\renewcommand{\d}{\delta}
\newcommand{\D}{\Delta}

\renewcommand{\th}{\theta}

\renewcommand{\l}{\lambda}

\newcommand{\s}{\sigma}

\renewcommand{\t}{\tau}

\renewcommand{\o}{\omega}

\newcommand{\<}{\langle}
\renewcommand{\>}{\rangle}
\newcommand{\ot}{\otimes}
\newcommand{\mod}{\mbox{mod}}
\renewcommand{\Re}{\mbox{Re}}
\renewcommand{\Im}{\mbox{Im}}

\newcommand{\sabd}{S_{ab}(\th-\th')}
\newcommand{\sab}{S_{ab}(\th)}

\newcommand{\rab}{\check{R}_{ab}(\th)}
\newcommand{\uqg}{U_q(\hat{g})}
\newcommand{\uqgt}{\widetilde{U_q(\hat{g})}}
\newcommand{\uqgu}{U_q(g^{(1)})}
\newcommand{\uqc}{U_q(c_n^{(1)})}
\newcommand{\antip}{{\cal S}}
\newcommand{\st}{s}

\newcommand{\htt}{\tilde{h}}
\newcommand{\Htt}{\tilde{H}}

\newcommand{\frap}{\cal F}
\newcommand{\dkC}{dual $(k)$-Coxeter }
\newcommand{\psa}{\pi_{s,a}}
\newcommand{\psat}{\pi_{s,a}^{(\th)}}
\newcommand{\psbt}{\pi_{s,b}^{(\th)}}

\begin{document}

\def\draft{\pagestyle{draft}\thispagestyle{draft}
\global\def\draftcontrol{1}}
\global\def\draftcontrol{0}

\newpage
\begin{titlepage}
\begin{flushright}
{KCL-TH-95-02}\\
{hep-th/9503079}
\end{flushright}
\vspace{2cm}
\begin{center}
{\bf {\large EXACT S-MATRICES WITH \\
\vspace{2mm}
AFFINE QUANTUM GROUP SYMMETRY}}\\
\vspace{1.5cm}
G.W. DELIUS\footnote{Supported by Habilitationsstipendium der
Deutschen Forschungsgesellschaft}
\footnote{On leave from Department of Physics, Bielefeld University,
Germany}\\
\vspace{2mm}
{\em Department of Mathematics}\\
{\em King's College London}\\
{\em Strand, London WC2R 2LS, UK}\\
{\small e-mail: delius@mth.kcl.ac.uk}\\
\vspace{1.6cm}
{\bf{ABSTRACT}}
\end{center}
\begin{quote}

We show how to construct the exact factorized S-matrices of
1+1 dimensional quantum field theories whose symmetry charges
generate a quantum affine algebra. Quantum affine Toda
theories are examples of such theories. We take into account
that the Lorentz spins of the symmetry charges determine the
gradation of the quantum affine algebras. This gives the
S-matrices a non-rigid pole structure. It dependson a kind of
``quantum'' dual Coxeter number which will therefore also
determine the quantum mass ratios in these theories.
As an example we explicitly construct S-matrices with
$U_q(c_n^{(1)})$ symmetry.

\end{quote}
\vfill
\end{titlepage}

\section{Introduction\lab{sectintro}}

This paper is concerned with the exact determination of the S-matrices
of certain 1+1 dimensional quantum field theories.

It is highly desirable to know S-matrices exactly because the complete
on-shell information about a quantum field theory is contained in its
S-matrix. In general however one is forced to resort to perturbative
or otherwise approximative solutions. Many interesting phenomena do
not show up in perturbation theory. The study of exact S-matrices
in 1+1 dimensions can shed light on such phenomena.
We would like to mention two examples: 1) The scalar S-matrices for the
fundamental particles in real coupling affine Toda theory
\ct{Bra90,Del92b} display a strong coupling --- weak coupling duality
which at the same time interchanges a Lie algebra with its dual algebra.
2) In the Sine-Gordon theory the exact breather S-matrix \ct{Zam79}
is equal to the S-matrix
of the fundamental particle, suggesting that the breather solutions and
fundamental particles are just different descriptions of the same object.
This has recently been extended to $a_2^{(1)}$ Toda theory \ct{Gan95}.
These are exactly the kind of phenomena which have been conjectured
to occur in 4-dimensional Yang-Mills theory \ct{God77,Mon77} and
have recently been followed up by Seiberg and Witten \ct{Sei94}.

The known exact 1+1 dimensional S-matrices are either diagonal or are
proportional to a rational or trigonometric R-matrix. Rational R-matrices
are intertwiners of representations of Yangian algebras \ct{Dri86} and
give the S-matrices of the principal chiral models \ct{Ber78,Ogi87}.
They have the feature that they do not depend on any parameter.
Trigonometric R-matrices are intertwiners of representations of
quantum affine algebras $U_q(\hat{g})$ \ct{Jim85,Dri86}. These
are deformations of the enveloping algebras of affine Kac-Moody algebras
\ct{Kac90} and depend on a parameter $q$. For $\hat{g}=a_n^{(1)}$ these
R-matrices give the soliton S-matrices of $a_n^{(1)}$ affine Toda
theory \ct{Hol93b} (and in particular for $a_1^{(1)}$ give the
Sine-Gordon S-matrix). This implies that these theories have a
$U_q(a_n^{(1)})$ quantum affine symmetry which had been observed
previously by Bernard and LeClair \ct{Ber91}. At $q$ a root of unity
the trigonometric R-matrices give the S-matrices of perturbed
W-invariant theories \ct{deV91,Bab94}.

In this paper we study quantum field theories with quantum
affine symmetry in general. We show how their S-matrices are obtained
from the universal R-matrices of the quantum affine algebras. Even
though the theory of quantum affine algebras and their quasitriangular
structure given by the universal R-matrices was not originaly
designed for this purpose, it turns out to be the ideal basis for
the construction of solutions to the S-matrix axioms. Indeed
the S-matrix axioms of unitarity, crossing symmetry
and the bootstrap principle follow directly from the fundamental
properties of the universal R-matrices. In addition, one gets
factorization, which expresses the multi-particle S-matrices in
terms of the two-particle S-matrices, for free. Schematically (we will
give details later):
\[
(\antip\ot 1)R=R^{-1}&\Rightarrow&\mbox{ crossing symmetry}\non
(\D\ot 1)R=R_{13}R_{12}&\Rightarrow&\mbox{ bootstrap principle}\\
R\D=\D^TR&\Rightarrow&\mbox{ unitarity and factorization}\nonumber
\]

We will show the importance of the
Lorentz spins of the symmetry charges. They
determine the gradation of the quantum affine algebras and the
S-matrices depends crucially on this. The influence which the gradation
has on the R-matrices was described by us in \ct{Bra94} and has
since been independently used in \ct{Bab94}. We show how the locations
of the particle poles of the S-matrices and therefore the particle
quantum mass ratios depend on both the deformation parameter $q$
and on the gradation, see eq.\Eq{poleloc}. In particular this
overcomes the restriction to unrenormalized mass ratios hitherto
observed in exact soltion S-matrices.

We study the requirement of crossing symmetry and find that it places
a constraint on the possible gradation. This leads to the formula
\Eq{poleloc2} in which a ``quantum'' dual Coxeter number occurs.
This extends the observation in real coupling Toda theory that
quantum effects tend to manifest themselves through the replacement
of the Coxeter number by a ``quantum'' Coxeter number \ct{Del92b}.

These general results are contained in section \ref{secttwo} and in
section \ref{sectcn} we demonstrate the general framework with the
example of S-matrices with $U_q(c_n^{(1)})$ quantum affine symmetry.

We will not include the application of our study to quantum affine
Toda theories in this paper. It will appear in a future joint publication with
Gerard Watts and Nial MacKay.

\section{S-matrices with quantum affine symmetry\lab{secttwo}}

Before specializing to an example in the next section, we will
here study the properties of the S-matrices of any two-dimensional
relativistic quantum field theory which has a quantum affine symmetry
$\uqg$. We start in section \ref{sectsymmetry} by defining what we mean by
a quantum affine symmetry. We introduce the two-particle S-matrix in
section \ref{sectsmatrix} and show how it is expressed through the R-matrix
of $\uqg$.
We give some relevant information
about the R-matrices in section \ref{sectrmatrix}.
In section \ref{sectcrossing} we derive unitarity and crossing
symmetry from the properties of $\uqg$.
We discuss the S-matrix pole structure in section \ref{sectbootstrap}.
In section \ref{sectspectrum} we will see how the quantum affine
symmetry fixes the particle pole locations and thus
determines the quantum mass ratios.

\subsection{Quantum affine symmetry\lab{sectsymmetry}}

We say that a relativistic quantum field theory has a quantum affine
symmetry $\uqg$ if the following two properties hold:

1) {\it The theory possesses quantum
conserved charges $H_i, X^\pm_i,~i=0\cdots r$, which obey the same relations
as the Chevalley generators of $\uqg$.}
Thus they obey the commutation relations
\[\label{crel}
\left[H_i,X^\pm_j\right]&=&\pm a_{ij} X^\pm_j,\nonumber\\
\left[X^+_i,X^-_j\right]&=&\delta_{ij}\frac{q_i^{H_i}-q_i^{-H_i}}
{q_i-q_i^{-1}},~~~~q_i\equiv q^{d_i},\\
\left[H_i,H_j\right]&=&[X^\pm_i,X^\pm_j]=0,\nonumber
\]
and also the quantum Serre relations, which we
will not write down here. For background on quantum affine
algebras see e.g. \ct{Cha94b}. In the above,
$a_{ij}$ is the generalized Cartan matrix of an affine Kac-Moody
algebra $\hat{g}$ \ct{Kac90} and the $d_i$ are coprime integers such that
the matrix $(d_i a_{ij})$ is symmetric.
$q$ is a complex parameter which will be related to
Planck's constant $\hbar$ and the coupling constant of the field theory.

2) {\it The conserved charges possess a definite Lorentz spin.}
Thus if $D$ denotes the infinitesimal two-dimensional Lorentz generator,
then we require that
\[\label{derivation}
[D,X^\pm_i]=\pm \st_i X^\pm_i,~~~~[D,H_i]=0,~~~~i=0,\dots,r.
\]
$\st_i\in\RR$ is called the Lorentz spin of $X^+_i$. The fact that the $X^-_i$
have Lorentz spin $-\st_i$ and that the $H_i$ have Lorentz spin $0$ is
required by consistency with the commutation relations \Eq{crel}.

The operators $X^\pm_i,H_i,~i=0,\dots,r,$ and $D$ together generate
the quantum enveloping algebra $\uqg$. $D$ is called the derivation.
It is because the Lorentz transformation is integrated into the
quantum affine symmetry algebra in this way, that this symmetry gives
strong constraints on the form of the S-matrix. We will denote the
algebra without the derivation $D$ by $\uqgt$.

Let the Lorentz spin of an operator $A$ be denoted by $\st(A)$. Then
$\st$ satisfies $\st(AB)=\st(A)+\st(B)$. Thus $\st:~\uqg \rightarrow \RR$ is a
gradation of $\uqg$. Such a gradation is uniquely fixed by the vector
$\st=(\st_0,\dots,\st_r)\in\RR^{r+1}$. The most common gradations used
in studying affine algebras are the homogeneous gradation which has
$s_0=1$ and all other $s_i=0$, and the principal gradation which has
all the $s_i=1$. We will see that interesting physical effects arise
from studying more general gradations, in particular gradations which
depend on the coupling constant of the field theory.

We now start to consider the consequences which the presence of a
quantum affine symmetry has for the theory.

The quantum affine symmetry implies quantum integrability of the
theory. Quantum integrability is given when one can find an infinite
number of commuting higher spin conserved charges. The infinitely
many Casimir operators
of $\uqgt$ supply such higher spin conserved charges.
They are not the standard local integer spin charges usually considered
\ct{Oli85},
but they have the same strong implications. For example their conservation
guarantees that in a scattering process the set of incoming momenta
equals the set of outgoing momenta. We arrange
the particles into multiplets under these charges. By a multiplet
we mean the collection of all particles with the same mass and the
same eigenvalues under all the higher spin Casimir operators.
The multiplets of one-particle states will transform in the finite
dimensional irreducible representations of $\uqgt$ uniquely determined
by the values of all the higher Casimir operators.

We denote the one-particle states by
$|a,\a,\th\rangle$, where $a$ denotes the multiplet, $\a$ labels the
particle within the multiplet and $\th$ is the rapidity of the particle.
The rapidity specifies the energy $E=m\, \mbox{cosh}(\th)$ and the momentum
$p=m\,\mbox{sinh}(\th)$ of the particle, $m$ being the mass of the particle.
At fixed rapidity the particles in the multiplet $a$ span the space
$V_a$ which carries a finite dimensional unitary
representation $\pi_a$ of $\uqgt$. The central charge of the algebra
takes the value zero in all finite dimensional
representations.\footnote{For the reader who is wondering how this is
consistent with
the statement that there are no finite dimenional unitary highest-weight
representations of affine Lie algebras: these are not highest-weight
representations in the usual sense. For a treatment of finite dimensional
representations of quantum affine algebras see \ct{Cha94a,Del94a}.}
Including the
rapidity the one-particle space is $V_a\otimes   \frap $, where
$\frap$ is a suitably chosen space of functions of $\th$.\footnote{Strictly
speaking the one-particle states of definite rapidity do not lie in this
space but need to be smeared by test functions as $\int\,d\th'\,f(\th-\th')
|a,\a,\th\>$, but all these details do not need to concern us here.}
Under a finite Lorentz transformation $L(\lambda)=\exp(\l D)$ the rapidity
$\th$ is shifted by $\l$
\[\label{lorentz}
L(\lambda)|a,\a,\th\rangle=|a,\a,\th+\l\rangle.
\]
{}From this we deduce that $V_a\ot\frap$ carries the following
infinite dimensional representation $\psa$ of $\uqg$, where the subscript
$s$ denotes the gradation
\[
\psa(D)&=&1\ot\frac{d}{d\th},\non
\psa(X_i^\pm)&=&\pi_a(X_i^\pm)\ot e^{\pm s_i\th},\non
\psa(H_i)&=&\pi_a(H_i)\ot 1.
\]
The appearance of $e^{\pm s_i\th}$ in $\psa$ is dictated by
\Eq{crel}.
Thus the one-particle states with definite rapidity $\th$ transform under
an element $A\in\uqgt$ as
\[
|a,\a,\th\rangle\mapsto
\psat(A)_{\a\b}|a,\b, \th\rangle
\]
where we have defined the family of finite dimensional representations
$\psat$ of $\uqgt$ by
\[
\psat(H_i)=\pi_a(H_i),~~~~\psat(X^\pm_i)=e^{\pm s_i\th}\pi_a(X^\pm_i).
\]
We will usually drop the subscript $s$ denoting the gradation if it is
clear from the context.

We can also derive the action of the symmetry on asymptotic
multi-particle states. We assume that
asymptotically, when the particles are far apart, a two-particle state can be
represented as a tensor product $|a,\a, \th\rangle\otimes|b,\b, \th'\rangle$
of two one-particle states. We choose the
ordering of the factors in the tensor product according to the ordering
of the particles in space, i.e. the first is to the left of the second.
Consistency with the commutation rules \Eq{crel} implies that the action
of the symmetry on such a state is given by the coproduct $\D$ of $\uqg$
\footnote{The algebra relations \Eq{crel} are invariant under
$q\leftrightarrow q^{-1}$, but our choice of the coproduct \ref{coproduct},
rather than its opposite, fixes $q$.}
\[\label{coproduct}
\D(H_i)&=&H_i\otimes 1+1\otimes H_i,~~~\D(D)=D\otimes 1+1\otimes D,\nonumber\\
\D(X^\pm_i)&=&X^\pm_i\otimes q_i^{H_i/2}+q_i^{-H_i/2}\otimes X^\pm_i.
\]
i.e that
\[
&&(|a,\a, \th\rangle\otimes|b,\b, \th'\rangle)\mapsto
\pi_{ab}^{ (\th\th')}(A)_{\a\a',\b\b'}
(|a,\a', \th\rangle\otimes|b,\b', \th'\rangle)\\
&&\mbox{where }\pi_{ab}^{(\th\th')}(A)=
\left((\pi_a^{(\th)})_{\a\a'}\otimes(\pi_b^{(\th')})_{\b\b'}\right)
\D(A),
\]
Such a nontrivial action on an asymptotic two-particle state, where the
action on the one particle depends on the state of the other particle even
though it is very far away, is possible only for nonlocal symmetry charges.
The action on n-particle states is
\[\label{multiact}
\pi_{a_1\cdots a_n}^{(\th_1\cdots\th_n)}(A)
=(\pi_{a_1}^{(\th_1)}\otimes\cdots\otimes\pi_{a_n}^{(\th_n)})
\D^{n-1}(A),
\]
where $\D^2=(1\otimes\D)\D,~~\D^3=(1\otimes 1\otimes\D)\D^2$, etc.

\subsection{The two-particle S-matrices}\lab{sectsmatrix}

We now introduce the two-particle S-matrices $\sabd$ describing
the process depicted in figure \ref{figsmatrix}.

\begin{figure}[htb]
\unitlength=1.00mm
\linethickness{0.4pt}
\begin{picture}(70.00,40.00)(-30,10)
\put(10.00,10.00){\vector(0,1){40.00}}
\put(10.00,10.00){\vector(1,0){60.00}}
\put(7.00,48.00){\makebox(0,0)[rc]{$t$}}
\put(67.00,8.00){\makebox(0,0)[ct]{$x$}}
\put(25.00,45.00){\line(1,-1){25.00}}
\put(25.00,20.00){\line(1,1){25.00}}
\put(25.00,18.00){\makebox(0,0)[ct]{$|a,\alpha,\theta\rangle$}}
\put(50.00,18.00){\makebox(0,0)[ct]{$|b,\beta,\theta'\rangle$}}
\put(50.00,46.00){\makebox(0,0)[cb]{$|a,\alpha',\theta\rangle$}}
\put(25.00,46.00){\makebox(0,0)[cb]{$|b,\beta',\theta'\rangle$}}
\put(38.00,32.00){\circle*{5.20}}
\put(46.00,32.00){\makebox(0,0)[lc]{$S_{ab}(\theta-\theta')$}}
\end{picture}
\caption{The two-particle scattering process described by $\sabd$
\lab{figsmatrix}}
\end{figure}
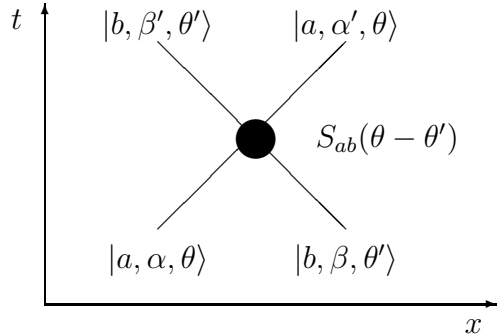
Note that due to the integrability of the theory, i.e. due to the
conservation of the higher-spin charges, these are the only two-particle
processes which are allowed. There can be no change of particle multiplets and
no change of rapidities. Only the particles within a multiplet can be
converted into each other. Lorentz invariance dictates that $S_{ab}$ depends
on $\th-\th'$ only. See \ct{Zam79} for a discussion of scattering in
integrable quantum field theories.

The S-matrix $ \sabd$ gives the mapping of an incoming asymptotic
two-particle state into an outgoing asymptotic two-particle state
\[
 \sabd:~V_a(\th)\otimes V_b(\th')\rightarrow
V_b(\th')\otimes V_a(\th)
\]
\[
|b,\b', \th'\rangle\otimes|a,\a', \th\rangle=
\left( \sabd\right)_{\a'\a,\b'\b}
\left(|a,\a, \th\rangle\otimes|b,\b, \th'\rangle\right)
\]
The quantum affine symmetry tells us that
\[\label{intertwine}
\sabd\pi_{ab}^{(\th\th')}(A)=\pi_{ba}^{(\th'\th)}(A)\sabd,
{}~~~~~\forall A\in\uqgt.
\]
This is just saying that, by the definition of a symmetry, it must not make
a difference whether we first perform a symmetry transformation and then
scatter or first scatter and then perform the symmetry transformation.

According to \Eq{intertwine} $\sabd$ is an intertwiner between the
representation $\pi_{ab}^{(\th\th')}$
and the representation $\pi_{ba}^{(\th'\th)}$. Because these representations
are irreducible for generic $\th$, $\th'$, such an intertwiner is unique, up to
an overall constant. This intertwiner is obtained by
evaluating the universal R-matrix of $\uqgt$ in the appropriate
representation and gradation
\[
\check{R}^{(s)}_{ab}(\th-\th')=\s_{ab}\left((\psat\otimes\psbt) R\right)
\]
and multiplying it by an overall scalar prefactor $f_{ab}$,
\[\label{sfromr}
\sabd=f_{ab}(\th-\th') \check{R}^{(s)}_{ab}(\th-\th').
\]
Here $\s_{ab}:~V_a(\th)\otimes V_b(\th')\rightarrow
V_b(\th')\otimes V_a(\th)$ is the permutation operator
$\s_{ab}: v_a\otimes v_b\mapsto v_b\otimes v_a$.
The prefactor $f_{ab}(\th)$ will be constrained by the requirements of
unitarity, crossing symmetry and the bootstrap principle, as we will
explain in the next sections.
That the right hand side of \Eq{sfromr} depends only on $\th-\th'$, as
required, follows from the fact that the universal R-matrix, as arising
from Drinfeld's double construction \ct{Dri86},
has the form $R=\sum_\c e_\c\otimes
e^\c$ with $s(e_\c)=-s(e^\c)$ for any gradation $s$.

By definition, the universal R-matrix of $\uqgt$ satisfies
\[\label{rdef}
R\D(A)=\D^T(A)R~~~~~\forall A\in\uqgt,
\]
where $\D^T$ is the opposite coproduct obtained by interchanging the
factors of the tensor products in \Eq{coproduct}. The intertwining
property \Eq{intertwine} of $\sabd$ follows immediately from \Eq{rdef}
by acting with $\s_{ab}(\pi_a^{(\th)}\otimes\pi_b^{(\th')})$ on both sides.

The multi-particle S-matrix needs to similarly intertwine the action
\Eq{multiact} of the symmetry on multi-particle states and from this
one can deduce that the multi-particle scattering is given as the product
of successive two-particle scatterings. The order in which the particles
interact pairwise is irrelevant, this is the Yang-Baxter equation.
Thus knowledge of the two-particle S-matrix is sufficient to describe
all scattering processes.

\subsection{R-matrices}\lab{sectrmatrix}

The matrices $\rab$ which enter the construction \Eq{sfromr} of the
S-matrices have a nice structure which we want to explain here.
Unfortunately the theory of quantum affine algebras has not yet been
developed as far as one would wish and we therefore have to restrict
our attention mainly to the untwisted algebras
$\uqgu$. Also we know details about $\rab$ only if the irreducible
$\uqgu$ representations $\pi_a$ and $\pi_b$ are both irreducible also as
representations of $U_q(g_0)$ and if furthermore the decomposition of
the tensor product representation $(\pi_a\otimes\pi_b)\D$ into
irreducible representations of $U_q(g_0)$ is multiplicity free. Here
$g_0$ denotes the finite dimensional Lie algebra associated to $g^{(1)}$.
When these conditions are not satisfied, only some isolated cases of
R-matrices have been determined but we hope that further developments
will take place soon. The general construction of the R-matrices
has been described for the case when these conditions hold in
\ct{Del94b}.

Let us start in the homogeneous gradation, i.e. the gradation
with $s_0=1$ and $s_i=0$ for $i=1,\dots,r$, which we will denote by
a super- or subscript $h$. In this gradation $\check{R}^{(h)}_{ab}(\th)\equiv
(\pi_{h,a}^{(\th)}\otimes\pi_{h,b}^{(0)})R$ takes the form
\[\label{formr}
\check{R}^{(h)}_{ab}(\th)=c_{ab}(\th)\sum_c\,\rho_{ab}^c(\th)\,
\check{P}^c_{ab}.
\]
The sum runs over the same values as the sum in the decomposition
$V_a\otimes V_b=\oplus_c V_c$
of the tensor product module $V_a\otimes V_b$ into irreducible
$U_q(g_0)$ modules $V_c$. The matrix $\check{P}^c_{ab}:~V_a\otimes V_b
\rightarrow V_c\subset V_b\otimes V_a$ is the $U_q(g_o)$ intertwiner
projecting onto $V_c$.
$\rho_{ab}^c(\th)$ is a function of the form
\[\label{rhos}
\rho_{ab}^c(\th)=\prod_{l\in L_{ab}^c}\langle l\rangle, ~~~~\mbox{where }
\langle l\rangle=\frac{1-e^\th q^l}{e^\th-q^l}.
\]
$L_{ab}^c$ is a set of integers. For details
see \ct{Del94b}. $c_{ab}(\th)$ is an overall scalar prefactor.
We exclude the case $q=1$ from our analysis because at
this point the $U_q(\hat{g})$-symmetric trigonometric R-matrices
collapse to rational R-matrices. Thus this case would have to be treated
separately.

Now we want to transfer these results for the homogeneous gradation to
a general gradation $s$ given by the Lorentz spins of the quantum
group generators. How to do this was explained in \ctz{section 5}{Bra94},
but we will repeat it here in our new notation. We note that we can
relate $\psat$ to $\pi_{h,a}^{(\mu\th)}$ for some $\mu$
as follows:
\[\label{dfg}
\pi_{h,a}^{(\mu\th)}\left(e^{\th H^{(s)}}X_i^\pm e^{-\th H^{(s)}}\right)=
e^{\pm\th\langle H^{(s)},\a_i\rangle \pm\th\mu\d_{i0}} \pi_a(X^\pm_i)=
\psat(X^\pm_i),
\]
provided the Cartan subalgebra element $H^{(s)}$ and the constant $\mu$
satisfy
\[
\langle H^{(s)},\a_i\rangle+{\mu}\d_{i0}=s_i,~~~~i=0,\dots,r.
\]
These equations fix $H^{(s)}$ and $\mu$. We can extract $\mu$ by multiplying
with the Kac labels $a_i$, summing over $i$ and using that $\sum_i
a_i \a_i=0$. We find that
\[
\mu=\sum_{i=0}^r \frac{a_i s_i}{a_0}.
\]
Relation \Eq{dfg} extends to the whole algebra
\[
\psat(A)=\pi_{h,a}^{(\mu\th)}\left(e^{\th H^{(s)}}A e^{-\th H^{(s)}}\right)
{}~~~\forall A\in\uqgt.
\]
This result allows us to relate the R-matrices
\[\label{rrelate}
R^{(s)}_{ab}(\th)&=&(\psat\otimes\pi_{s,b}^{(0)})R=
(\pi_{h,a}^{(\mu\th)}\otimes\pi_{h,b}^{(0)})
\left(\left(e^{\th H^{(s)}}\otimes 1\right)R\left(e^{-\th H^{(s)}}
\otimes 1\right)\right)
\nonumber\\
&=&\left(\pi_a(e^{\th H^{(s)}})\otimes 1\right)R_{ab}^{(h)}(\mu\th)
\left(\pi_a(e^{-\th H^{(s)}})\otimes 1\right),
\]
and thus, finally,
\[\label{formrs}
\check{R}^{(s)}_{ab}(\th)&=&c_{ab}(\mu\th)\sum_c\rho^c_{ab}(\mu\th)
\check{P}^{(s)c}_{ab}\nonumber\\
\mbox{with}&&\check{P}^{(s)c}_{ab}=
\left(1\otimes\pi_a(e^{\th H^{(s)}})\right)
\check{P}^c_{ab}\left(\pi_a(e^{-\th H^{(s)}})\otimes1\right).
\]

\subsection{Unitarity and crossing symmetry}\lab{sectcrossing}

In S-matrix theory one analytically extends the S-matrix $S(\th)$ to
complex values of the rapidity.
This analytic S-matrix for a relativistic quantum field theory has to
satisfy certain requirements. It has to be unitary
\[\label{unitarity}
S_{ba}(-\th)\cdot \sab=1\otimes 1
\]
and crossing symmetric
\[\label{crossing1}
\sab&=&(C_b^{-1}\otimes
1)\left(\s_{a\bar{b}}S_{\bar{b}a}(i\pi-\th)\right)^{t_1}
\s_{a\bar{b}}\,(1\otimes C_b)\\
&=&(1\otimes C_a^{-1})
\left(\s_{\bar{a}b} S_{b\bar{a}}(i\pi-\th)\right)^{t_2}
\s_{\bar{a}b}\,(C_a\otimes 1).
\label{crossing2}
\]
Here $C_a$ is the charge conjugation
matrix mapping a particle into its anti-particle. The superscript $t_1$
denotes transposition in the first factor, $t_2$ transposition in the
second factor.
These relations are easier to understand
when seen graphically as in figure \ref{figcrossing}.
See e.g. \ct{Zam79} for a treatment of S-matrices in 1+1 dimensions.

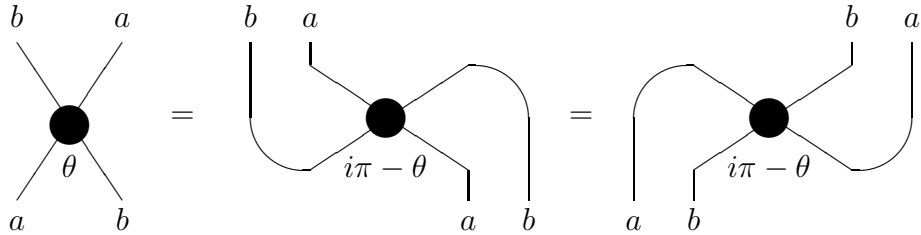
\begin{figure}[htb]
\unitlength=1.00mm
\linethickness{0.4pt}
\begin{picture}(131.00,31.00)(-5,25)
\put(12.00,50.00){\line(2,-3){14.00}}
\put(12.00,29.00){\line(2,3){14.00}}
\put(51.00,47.00){\line(3,-2){21.00}}
\put(51.00,33.00){\line(3,2){21.00}}
\put(19.00,39.00){\circle*{5.20}}
\put(61.00,40.00){\circle*{5.20}}
\put(19.00,35.00){\makebox(0,0)[ct]{$\theta$}}
\put(61.00,35.00){\makebox(0,0)[ct]{$i\pi-\theta$}}
\put(72.00,33.00){\line(0,-1){4.00}}
\put(72.00,27.00){\line(0,0){0.00}}
\put(51.00,47.00){\line(0,1){3.00}}
\put(51.00,41.00){\oval(16.00,16.00)[lb]}
\put(43.00,41.00){\line(0,1){9.00}}
\put(72.00,39.00){\oval(16.00,16.00)[rt]}
\put(80.00,39.00){\line(0,-1){10.00}}
\put(102.00,47.00){\line(3,-2){21.00}}
\put(102.00,33.00){\line(3,2){21.00}}
\put(112.00,40.00){\circle*{5.20}}
\put(112.00,35.00){\makebox(0,0)[ct]{$i\pi-\theta$}}
\put(102.00,39.00){\oval(16.00,16.00)[lt]}
\put(123.00,41.00){\oval(16.00,16.00)[rb]}
\put(131.00,41.00){\line(0,1){9.00}}
\put(94.00,39.00){\line(0,-1){10.00}}
\put(102.00,29.00){\line(0,1){4.00}}
\put(123.00,47.00){\line(0,1){3.00}}
\put(87.00,40.00){\makebox(0,0)[cc]{$=$}}
\put(34.00,40.00){\makebox(0,0)[cc]{$=$}}
\put(12.00,25.00){\makebox(0,0)[cb]{$a$}}
\put(26.00,25.00){\makebox(0,0)[cb]{$b$}}
\put(72.00,25.00){\makebox(0,0)[cb]{$a$}}
\put(80.00,25.00){\makebox(0,0)[cb]{$b$}}
\put(94.00,25.00){\makebox(0,0)[cb]{$a$}}
\put(102.00,25.00){\makebox(0,0)[cb]{$b$}}
\put(12.00,52.00){\makebox(0,0)[cb]{$b$}}
\put(26.00,52.00){\makebox(0,0)[cb]{$a$}}
\put(43.00,52.00){\makebox(0,0)[cb]{$b$}}
\put(51.00,52.00){\makebox(0,0)[cb]{$a$}}
\put(123.00,52.00){\makebox(0,0)[cb]{$b$}}
\put(131.00,52.00){\makebox(0,0)[cb]{$a$}}
\end{picture}
\caption{Crossing symmetry relations\lab{figcrossing}}
\end{figure}

Unitarity \Eq{unitarity} can always be ensured for the S-matrix
defined in \Eq{sfromr}. To see this we set $\th'=0$ and multiply
\Eq{intertwine} by $S_{ba}(-\th)$ from the left
\[\label{derunitarity}
S_{ba}(-\th)\sab\pi_{ab}^{(\th\th')}(A)&=&
S_{ba}(-\th)\pi_{ba}^{(\th'\th)}(A)\sab\nonumber\\
&=&\pi_{ab}^{(\th\th')}(A)S_{ba}(-\th)\sab~~~~
\forall A\in\uqgt.
\]
Because the representation $\pi_{ab}^{(\th\th')}$ is irreducible for generic
$\th,\th'$, \Eq{derunitarity} implies by Schurr's lemma that
$S_{ba}(-\th)\sab$ is proportional to the identity. We can
of course also derive this from the explicit expression \Eq{formr}
for $\check{R}_{ab}$ and find
\[\label{arf}
\check{R}_{ba}(-\th)\rab=c_{ba}(-\th)c_{ab}(\th)(1\otimes 1).
\]
Thus $\sab$ will satisfy \Eq{unitarity} if we choose $f_{ab}$ to satisfy
\[
f_{ba}(-\th)f_{ab}(\th)=c^{-1}_{ba}(-\th)c^{-1}_{ab}(\th).
\]
We would like to stress that the fact that the S-matrix is unitary
is not related to the question wether the corresponding field theory
is unitary. Even non-unitary theories have unitary S-matrices.

Proving that $\sab$ satisfies the crossing symmetry relations \Eq{crossing1}
and \Eq{crossing2} is a little more involved.
We will need the relation between the antipode and the charge conjugation
matrix. Let us first review the classical case of a Lie algebra $g$.
To each irreducible representation $\pi_a$ of $g$ acting on a representation
space $V_a$ one defines the dual representation $\pi_a^*$ acting on
$V_a^*$ by $\pi_a^*(A)=\pi_a(\c(A))^t~\forall A\in g$, where $\c$ is the
antipode of $g$ (which simply acts as $\c(A)=-A~\forall A\in g$).
The superscript $t$ denotes transposition of the representation matrix.
The classical charge
conjugation matrix $C_a^{cl}$ is the symmetric matrix defined by
\[\label{cldual}
\pi_a^*(A)^t\equiv\pi_a\left(\c(A)\right)=
\left(C_a^{cl}\right)^{-1}\pi_{\bar{a}}(A)^t\,C_a^{cl},
\]
where $\pi_{\bar{a}}$ is the conjugate representation to $\pi_a$
which is usually $\pi_a$ itself but is sometimes only related to it by
a diagram automorphism $\t$ as
$\pi_{\bar{a}}(A)=\pi_a\left(\tau(A)\right)$.

For quantum groups $U_q(g)$ based on a finite dimensional Lie algebra $g$
the situation is only changed slightly. The antipode $\antip$ of $U_q(g)$ acts
as
\[\label{antipode}
\antip(A)=q^{H_\rho}\,\c(A)\,q^{-H_\rho},
\]
where $\langle H_\rho,\a_i\rangle=d_i$ for all simple roots $\a_i$.
Thus the charge conjugation matrix is changed to
\[
C_a^{fin}=C_a^{cl}\,\pi_a(q^{-H_\rho}).
\]

For the representations $\pi_a^{(\th)}$ of quantum affine algebras $\uqgt$
the situation is  more interesting because
in addition to conjugation by a charge conjugation matrix also the spectral
parameter $\th$ needs to be shifted. This is so because if we
write the antipode of $\uqg$ as in \Eq{antipode}, then $H_\rho$
contains a component in the direction of the derivation $D$, i.e. the
Lorentz boost generator. Alternatively we can use
$\antip(X^\pm_i)=-q^{\pm d_i}\,X^\pm_i$ and write
\[
\pi_a^{(\th)}\left(\antip(X^\pm_i)\right)=
-q^{\pm d_i}\,e^{\pm \th s_i}\,\pi_a(X^\pm_i)=
\left(e^{\pm(\th+\xi)s_i}\right)\,
\pi_a\left(-q^{H_\chi}\,X^\pm_i\,q^{-H_\chi}\right),
\]
where the Cartan subalgebra element
$H_\chi$ has no component in the direction of the derivation.
The last equality holds if $H_\chi$ and $\xi$ satisfy
\[\label{preeq}
e^{s_i\xi} q^{\< H_{\chi},\a_i\>}=q^{d_i}
\]
If $q$ is a real number, then this determines
$H_\chi$ and $\xi$. We want to treat also
\footnote{In fact we will be interested later only in the case where
$q$ is a pure phase, so that the poles of the R-matrix appear for
purely imaginary $\th$.} the case of
complex $q$ and then there is a freedom due to the $2\pi i$ ambiguity
in the phases. Therefore we write $q=\exp{\o}$ with $\o$ a complex
number and introduce the notation
\[
[x]\equiv\Re(x)+i\left(\Im(x)\mod 2\pi\right)
\]
Then \Eq{preeq} is equivalent to
\[\label{eqforchi}
[\o]\langle H_\chi,\a_i\rangle +\xi s_i=[\o] d_i+2\pi i m_i,
{}~~~m_i\in\ZZ~\mbox{arbitrary}.
\]
My realization that there is this arbitrariness can be traced back
to a discussion with Nial MacKay.
We are most interested in $\xi$.
We extract it from \Eq{eqforchi} by multiplying with the Kac labels
$a_i$, summing over $i$ and using that $\sum_{i=0}^r\,a_i\a_i=0$. We
obtain
\[\label{cco}
\xi=\frac{[\o]}{a_0 \mu} \htt +\frac{2\pi i m}{a_0\mu},~~~
m\in\ZZ~\mbox{arbitrary},
\]
where
\[
\mu=\sum_{i=0}^r \frac{a_i s_i}{a_0},~~~~~
\htt=\sum_{i=0}^r a_i d_i.
\]
We see that $\mu$ is the same as in section \ref{sectrmatrix}.
$\htt$ is the \dkC number. It can be expressed in terms of the
dual Coxeter number $h^\vee$ and the twist $k^\vee$ of the dual algebra
as $\htt=k^\vee h^\vee$.
We will only consider
the case where $s_i=s_{\tau(i)}$. We arrive at the generalization
of \Eq{cldual} to the quantum affine case
\[\label{quas}
\pi_a^{(\th)}\left(\antip(A)\right)=
C_a^{-1}\,\pi_{\bar{a}}^{(\th+\xi)}(A)^t\,C_a,~~~~\forall A\in\uqgt
\]
where
\[
C_a=C_a^{cl}\,\pi_a(q^{-H_\chi}).
\]
In the special case of the homogeneous gradation
the shift in the spectral parameter had already been observed by
Frenkel and Reshetikhin in \ct{Fre92}\footnote{What they call the dual
Coxeter number is however really the dual $(k)$-Coxeter number
$\tilde{h}=k^\vee h^\vee$.}.

We are now ready to derive the crossing relations from the following
properties of the universal R-matrix
\[
(\antip\otimes 1)\,R=R^{-1},~~~(1\otimes \antip^{-1})\,R=R^{-1}.
\]
We will show how to derive \Eq{crossing1} from the first of these
equations, the derivation of \Eq{crossing2} from the second is
analogous. Acting with $\pi_b^{(0)}\otimes\pi_a^{(\th)}$ on both sides of the
equation and using \Eq{quas} and \Eq{arf} we find
\[
(C_b^{-1}\otimes 1)\left(R_{\bar{b}a}(\xi-\th)\right)^{t_1}
(C_b\otimes 1)=R^{-1}_{ba}(-\th)=
c_{ba}^{-1}(-\th)c_{ab}^{-1}(\th)
\s_{ab}R_{ab}(\th)\s_{ba},
\]
where we have defined $(\pi_a^{(\th)}\otimes\pi_b^{(\th')})R=
R_{ab}(\th-\th')$.
Using \Eq{sfromr} this can be rewritten in terms of the S-matrix
\[
&&(C_b^{-1}\otimes 1)
\left(\s_{a\bar{b}}S_{\bar{b}a}(\xi-\th)\right)^{t_1}
\s_{a\bar{b}}(1\otimes C_b)\nonumber\\
&&=c_{ba}^{-1}(-\th)c_{ab}^{-1}(\th)
f_{\bar{b}a}(\xi-\th)f^{-1}_{ab}(\th)S_{ab}(\th).
\]
This produces \Eq{crossing1} if
\[
\xi=i\pi~~\mbox{and}~~
f_{ab}(i\pi-\th)=f_{ab}^{-2}(\th)f_{ba}(-\th),
\]
thus putting further strong constraints on the scalar prefactor $f_{ab}$.
We observe that crossing symmetry places a constraint on the possible
gradations
\[
i\pi=\xi=\frac{[\o]}{a_0 \mu} \htt +\frac{2\pi i m}{a_0\mu},~~~
m\in\ZZ~\mbox{arbitrary}.
\]
Thus only gradations $(s)$ for which
\[\label{cc}
a_0\mu\equiv\sum_{i=0}^r a_i s_i=
\frac{[\o]}{i\pi}\htt+2m,~~~m\in\ZZ
\]
can lead to crossing symmetric S-matrices.

\subsection{S-matrix poles}\lab{sectbootstrap}

A lot of information is contained in the pole structure of the analytic
S-matrix. Indeed, the whole mass spectrum and the three-particle fusion
rules can be read off from the location of the simple poles. Conversely,
knowledge of the spectrum and the three-particle couplings determines
the pole structure of the S-matrices. Because the S-matrices of integrable
quantum field theories have to also obey other constraints, in particular
the bootstrap principle, only very particular kinds of spectra can be
realized in these theories.

If particles of type $a$ and $b$ can form a bound state of type $c$ with
mass
\[\label{massc}
m_c^2=m_a^2+m_b^2+2m_a m_b \cos u_{ab}^c,
\]
then this usually leads to a simple pole of $S_{ab}(\th)$ at
$\th=i u_{ab}^c$ corresponding to the propagation of a particle $c$
in the direct channel, as depicted in figure \ref{figpoles} a).
By crossing symmetry it also leads to poles in $S_{b\bar{a}}$ and
$S_{\bar{b}a}$ at $\th=i\bar{u}_{ab}^c=i(\pi-u_{ab}^c)$, see figure
\ref{figpoles} b) and c).

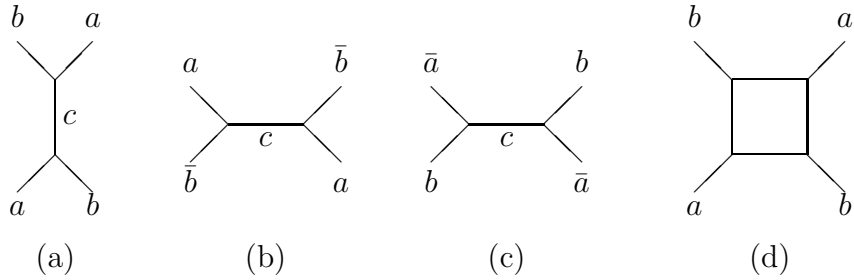
\begin{figure}[htb]
\unitlength=1.00mm
\linethickness{0.4pt}
\begin{picture}(125.00,37.00)(-5,10)
\put(20.00,40.00){\line(0,-1){10.00}}
\put(20.00,30.00){\line(-1,-1){5.00}}
\put(20.00,30.00){\line(1,-1){5.00}}
\put(20.00,40.00){\line(-1,1){5.00}}
\put(20.00,40.00){\line(1,1){5.00}}
\put(75.00,34.00){\line(1,0){10.00}}
\put(85.00,34.00){\line(1,1){5.00}}
\put(85.00,34.00){\line(1,-1){5.00}}
\put(75.00,34.00){\line(-1,1){5.00}}
\put(75.00,34.00){\line(-1,-1){5.00}}
\put(15.00,22.00){\makebox(0,0)[cb]{$a$}}
\put(25.00,22.00){\makebox(0,0)[cb]{$b$}}
\put(21.00,35.00){\makebox(0,0)[lc]{$c$}}
\put(15.00,47.00){\makebox(0,0)[cb]{$b$}}
\put(25.00,47.00){\makebox(0,0)[cb]{$a$}}
\put(70.00,25.00){\makebox(0,0)[cb]{$b$}}
\put(90.00,25.00){\makebox(0,0)[cb]{$\bar{a}$}}
\put(70.00,41.00){\makebox(0,0)[cb]{$\bar{a}$}}
\put(90.00,41.00){\makebox(0,0)[cb]{$b$}}
\put(43.00,34.00){\line(1,0){10.00}}
\put(53.00,34.00){\line(1,1){5.00}}
\put(53.00,34.00){\line(1,-1){5.00}}
\put(43.00,34.00){\line(-1,1){5.00}}
\put(43.00,34.00){\line(-1,-1){5.00}}
\put(38.00,25.00){\makebox(0,0)[cb]{$\bar{b}$}}
\put(58.00,25.00){\makebox(0,0)[cb]{$a$}}
\put(38.00,41.00){\makebox(0,0)[cb]{$a$}}
\put(58.00,41.00){\makebox(0,0)[cb]{$\bar{b}$}}
\put(48.00,33.00){\makebox(0,0)[ct]{$c$}}
\put(80.00,33.00){\makebox(0,0)[ct]{$c$}}
\put(110.00,40.00){\line(0,-1){10.00}}
\put(110.00,30.00){\line(1,0){10.00}}
\put(120.00,30.00){\line(0,1){10.00}}
\put(120.00,40.00){\line(-1,0){10.00}}
\put(110.00,40.00){\line(-1,1){5.00}}
\put(120.00,40.00){\line(1,1){5.00}}
\put(110.00,30.00){\line(-1,-1){5.00}}
\put(120.00,30.00){\line(1,-1){5.00}}
\put(105.00,22.00){\makebox(0,0)[cb]{$a$}}
\put(125.00,22.00){\makebox(0,0)[cb]{$b$}}
\put(125.00,47.00){\makebox(0,0)[cb]{$a$}}
\put(105.00,47.00){\makebox(0,0)[cb]{$b$}}
\put(20.00,15.00){\makebox(0,0)[cb]{(a)}}
\put(48.00,15.00){\makebox(0,0)[cb]{(b)}}
\put(80.00,15.00){\makebox(0,0)[cb]{(c)}}
\put(115.00,15.00){\makebox(0,0)[cb]{(d)}}
\end{picture}

\caption{Processes leading to poles in the S-matrix
\lab{figpoles}}
\end{figure}

Higher order poles in the S-matrix are due to higher order processes like
the one depicted in figure \ref{figpoles} d). When and where such processes
give higher order poles according to the Coleman-Thun mechanism
is well explained in \ct{Bra90,Bra91a}.

We say that a bound state with a mass $m_c$ given by \Eq{massc}
only ``usually'' leads to a simple pole at $\th=i u_{ab}^c$ because
it has been observed \ct{Del92a,Del92b} that higher order processes like
that in figure
\ref{figpoles} d) might take place at a value of the rapidity which is very
close to $i u_{ab}^c$. In that case the outcome may be a simple pole slightly
shifted away from $i u_{ab}^c$. This phenomenon is important in affine
Toda theory.

The residue of a particle pole at $\th=i u_{ab}^c$ of the S-matrix
$S_{ab}(\th):\,V_a\otimes V_b\rightarrow V_b\otimes V_a$
should project onto an irreducible submodule $V_c$, i.e., $\mbox{Res}
(S_{ab}(\th=i u_{ab}^c)):
\,V_a\otimes V_b\rightarrow V_c\subset V_b\otimes V_a$. This is just
saying that the scattering process at the pole is dominated by the
propagation of particles of type $c$ only. Our S-matrix
$S_{ab}(\th)=f_{ab}(\th)\check{R}_{ab}^{(s)}(\th)$ has the potential of
producing such a phenomenon. Looking at \Eq{formrs} we see that at any value
of $\th$ at which some of the $\rho_{ab}^d(\mu\th)$ have a pole, the
residue of
$\check{R}_{ab}^{(s)}(\th)$ indeed projects onto a submodule. Unless the
scalar prefactor had a zero at the same location, the particles
$a$ and $b$ would form a bound state transforming in the corresponding
sub-representation. If only one of the $\rho_{ab}^d$, let us say
$\rho_{ab}^c$, has a pole at  $\th=i u_{ab}^c$ then the bound state
particle transforms in the irreducible representation $\pi_c$ of
$U_q(g)$. In general however, several of the $\rho_{ab}^d$, let us say
$\rho_{ab}^{c_1},\dots,\rho_{ab}^{c_n}$, will have a pole, in which case the
bound state particle transforms as the representation
$\pi_c=\pi_{c_1}\oplus\cdots\oplus\pi_{c_n}$ which is reducible as a
representation of $U_q(g)$ but still irreducible as a representation of
$\uqg$.

If a theory has a particle spectrum containing the multiplets
$a_1,\dots,a_n$, corresponding to some irreducible $\uqg$ modules
$V_{a_1},\dots,V_{a_n}$, then one has to ensure that the corresponding
S-matrices $S_{a_i a_j}$ have particle poles only at locations such that
the residue projects onto one of the modules $V_{a_1},\dots,V_{a_n}$.
If there was a simple pole with a residue projecting onto some other
module $V_b$ then a corresponding particle multiplet $b$ would also
have to be in the spectrum of the theory. If there was a simple pole\
at a location where the residue is not a projector onto a submodule,
then that pole must be checked to have an explanation in terms of
the Coleman-Thun mechanism. For examples of such poles see
\ct{Del92b},\ct{Cor93}.

We realize from these comments that one should choose the prefactor $f_{ab}$
in such a way that it cancels many of the poles in the expression \Eq{formr}
for the $\check{R}$-matrix which would lead to bound states which do not
actually exist in the theory. That doing this is a difficult task is due to
the bootstrap principle.

\begin{figure}[htb]
\unitlength=1mm
\linethickness{0.4pt}
\begin{picture}(96.00,42.00)(-25,12)
\put(73.00,20.00){\line(2,3){10.00}}
\put(83.00,35.00){\line(2,-3){10.00}}
\put(83.00,35.00){\line(0,1){15.00}}
\put(20.00,20.00){\line(2,3){10.00}}
\put(30.00,35.00){\line(2,-3){10.00}}
\put(30.00,35.00){\line(0,1){15.00}}
\put(67.00,20.00){\line(5,3){28.00}}
\put(12.00,33.00){\line(5,3){28.00}}
\put(30.00,44.00){\circle*{2.00}}
\put(77.00,26.00){\circle*{2.00}}
\put(86.00,31.00){\circle*{2.00}}
\put(20.00,16.00){\makebox(0,0)[cb]{$a$}}
\put(40.00,16.00){\makebox(0,0)[cb]{$b$}}
\put(67.00,16.00){\makebox(0,0)[cb]{$d$}}
\put(73.00,16.00){\makebox(0,0)[cb]{$a$}}
\put(93.00,16.00){\makebox(0,0)[cb]{$b$}}
\put(30.00,52.00){\makebox(0,0)[cb]{$c$}}
\put(40.00,52.00){\makebox(0,0)[cb]{$d$}}
\put(83.00,52.00){\makebox(0,0)[cb]{$c$}}
\put(11.00,29.00){\makebox(0,0)[cb]{$d$}}
\put(96.00,39.00){\makebox(0,0)[cb]{$d$}}
\put(53.00,35.00){\makebox(0,0)[cc]{$=$}}
\end{picture}
\caption{The bootstrap principle\lab{figbootstrap}}
\end{figure}

The bootstrap principle of S-matrix theory states that if $S_{ab}(\th)$
has a simple pole corresponding to a particle of type $c$, then the
S-matrices $S_{dc}$ describing the scattering of a particle type $c$ with
any other particle type $d$ are expressed in terms of the S-matrices
$S_{ad}$ and $S_{bd}$. This is expressed pictorially in figure
\ref{figbootstrap} and through the formula
\[\label{bootstrap}
S_{dc}(\th)\left(1\otimes P_{ab}^c\right)=
\left(1\otimes S_{db}(\th+i \bar{u}_{bc}^a)\right)
\left(S_{da}(\th-i\bar{u}_{ac}^b)\otimes 1\right),
\]
where $P_{ab}^c$ is the projector onto $V_c$ in $V_a\otimes V_b$.
That the matrices $\rab$ satisfy such a relation follows from the
defining property of the universal R-matrix \ct{Dri86}
\[
(1\otimes \D)R=R_{13}R_{12}.
\]
That also $\sab$ satisfies \Eq{bootstrap} puts further constraints
on the scalar prefactor $f_{ab}$ and in particular on the location
of its poles and zeros.

\subsection{General remarks on the quantum spectrum\lab{sectspectrum}}

The simple particle poles of the S-matrix will be at locations
at which the R-matrix projects onto a submodule. This implies that
they occur at values of $\th$ at which
$e^{\mu\th}=q^{l}$, where $l$ is one of the numbers in eq. \Eq{rhos}.
Writing again $q=e^\o$ we see that the potential particle poles occur at
\[\label{poleloc}
\th=l\frac{[\o]}{\mu}+\frac{2\pi i p}{\mu},~~~p\in\ZZ.
\]
At which of these potential locations the S-matrix will really
have poles is of course determined by the zeros of the scalar prefactor.
Clearly we will have to have $\o$ purely imaginary (i.e. $q$ a pure
phase) in order for the poles to lie on the imaginary axis, as is
required for stable particle poles. $\mu$ is real by construction
because the Lorentz spins of the symmetry charges are real.

Let us assume that we know the classical spectrum of particles and
their coupling rules in
the integrable field theory under study. Then we know the classical
locations of the poles. We can identify them among the poles in
\Eq{poleloc} and then we can read off their dependence of $\hbar$
and the coupling constant from the dependence of $[\o]$ and $\mu$
on these. By this procedure we can derive the full quantum mass
ratios of the particles. (When we say ``particle'' we mean of course
not only fundamental particles but also solitons, breathers, excited
solitons etc.).

This is of great significance, because it is usually
next to impossible to calculate the quantum corrections
to masses to all orders. Usually, even calculating just the first
order correction to the masses is a formidable task, as evidenced
by recent calculations of the first mass corrections to the soliton
masses in affine Toda theory \ct{Hol93a,Del94c,Mac94}. On the other hand
it is usually simpler to determine the existence of symmetry algebras
to all orders, because here one can often make use of the fact that
no further anomalies can appear beyond a certain orders in perturbation
theory. See for example the proof of quantum integrability of real
coupling Toda theory \ct{Del92c}. Similarly Bernard and LeClair have
argued the quantum affine symmetry in imaginary coupling Toda
theory to all orders by a scaling argument \ct{Ber91}.

{}From the freedom of choosing the integer $p$ in \Eq{poleloc} we
see that to any particle transforming in a particular representation
$c$ there can be further particle
states transforming in the same representation, corresponding to other
values of $p$. These could be interpreted as excitations of the particle.
Because particle poles have to lie on the physical sheet, i.e.
at $0<\Im \th<i \pi$, these states can exist only for the integers
$p$ in a certain range. This range is determined by the gradation
through the parameter $\mu$. Conversely, if one does not know the
gradation, but knows  the tower of excited states in the spectrum,
then one can deduce $\mu$ by the separation between these states
from \Eq{poleloc}.

It is illuminating to rewrite the pole locations in \Eq{poleloc}
using the constraint \Eq{cc} coming from crossing symmetry. We
obtain
\[\label{poleloc2}
\th=l a_0\frac{i\pi}{\Htt}+
\frac{2\pi i p}{\mu}.
\]
where we have introduced a sort of ``quantum'' \dkC number
\[
\Htt=\htt+\frac{2\pi i m}{[\o]}
\]
This is very reminiscent of the ``quantum'' Coxeter number $H$
which appears in the pole locations of the scalar S-matrices for
the fundamental particles in real coupling affine Toda theory
\ct{Del92a,Del92b}.

\section{$\uqc$ symmetric S-matrices}\lab{sectcn}

In this section we will give a concrete example for a consistent
set of S-matrices for a theory with the quantum affine symmetry
based on the Kac-Moody algebra $c_n^{(1)}$.

To specify a theory we have to not only give the symmetry algebra
but also state which representations of the symmetry algebra will
occur as particle multiplets. In this example we choose to include
particles transforming in all the fundamental representations of
$\widetilde{\uqc}$. The fundamental representations are the
representations whose highest weight is a fundamental weight $\l_a$ of
$c_n$. The fundamental weights are defined by the property that
$\l_a\cdot\a^\vee_b=\d_{ab},~a,b=1,\dots,r$.

We have two related reasons for choosing this particular example.

1) There is a lagrangian field theory which exhibits exactly these
features: $d_{n+1}^{(2)}$ Toda theory at imaginary coupling
is believed to have a $c_n^{(1)}$
quantum affine symmetry \ct{Ber91} and its solitons species
\ct{Oli93b}
correspond to the fundamental representations. It is a long
standing problem to construct the corresponding soliton S-matrices
and it is hoped that this example will lead to the solution
of that problem.

2) Hollowood \ct{Hol94} has already attempted to construct S-matrices
of this type.
He finds it to be impossible to give a suitable scalar prefactor
to implement the correct pole structure. We can now see that this
failure is due to the fact that he implicitly worked with a gradation
with $\mu=[\o]\htt$. Using a more generic gradation, the construction
becomes possible.

\subsection{The R-matrices}

\begin{figure}[ht]
\unitlength=1.25mm
\linethickness{0.4pt}
\begin{picture}(120.00,70.00)(17,3)
\put(30.00,10.00){\circle*{4.00}}
\put(46.00,10.00){\circle*{4.00}}
\put(38.00,18.00){\circle*{4.00}}
\put(54.00,18.00){\circle*{4.00}}
\put(130.00,10.00){\circle*{4.00}}
\put(114.00,10.00){\circle*{4.00}}
\put(122.00,18.00){\circle*{4.00}}
\put(62.00,10.00){\circle*{4.00}}
\put(46.00,26.00){\circle*{4.00}}
\put(80.00,60.00){\circle*{4.00}}
\put(72.00,52.00){\circle*{4.00}}
\put(88.00,52.00){\circle*{4.00}}
\put(96.00,44.00){\circle*{4.00}}
\put(80.00,44.00){\circle*{4.00}}
\put(30.00,10.00){\line(1,1){20.00}}
\put(46.00,10.00){\line(1,1){12.00}}
\put(46.00,26.00){\line(1,-1){16.00}}
\put(38.00,18.00){\line(1,-1){8.00}}
\put(80.00,60.00){\line(-1,-1){12.00}}
\put(80.00,60.00){\line(1,-1){20.00}}
\put(72.00,52.00){\line(1,-1){12.00}}
\put(88.00,52.00){\line(-1,-1){12.00}}
\put(130.00,10.00){\line(-1,1){12.00}}
\put(122.00,18.00){\line(-1,-1){8.00}}
\put(114.00,10.00){\line(-1,1){4.00}}
\put(96.00,44.00){\line(-1,-1){4.00}}
\put(62.00,10.00){\line(1,1){4.00}}
\put(80.00,65.00){\makebox(0,0)[cb]{$\l_a$}}
\put(88.00,65.00){\makebox(0,0)[cb]{$\l_{a+1}$}}
\put(96.00,65.00){\makebox(0,0)[cb]{$\l_{a+2}$}}
\put(105.00,65.00){\makebox(0,0)[cb]{$\cdots$}}
\put(122.00,65.00){\makebox(0,0)[cb]{$\l_{a+b-1}$}}
\put(130.00,65.00){\makebox(0,0)[cb]{$\l_{a+b}$}}
\put(73.00,65.00){\makebox(0,0)[cb]{$\l_{a-1}$}}
\put(30.00,65.00){\makebox(0,0)[cb]{$\l_{a-b}$}}
\put(38.00,65.00){\makebox(0,0)[cb]{$\l_{a-b+1}$}}
\put(59.00,65.00){\makebox(0,0)[cb]{$\cdots$}}
\put(25.00,60.00){\makebox(0,0)[rc]{$\l_b$}}
\put(25.00,52.00){\makebox(0,0)[rc]{$\l_{b-1}$}}
\put(25.00,44.00){\makebox(0,0)[rc]{$\l_{b-2}$}}
\put(25.00,36.00){\makebox(0,0)[rc]{$\vdots$}}
\put(25.00,10.00){\makebox(0,0)[rc]{$0$}}
\put(25.00,18.00){\makebox(0,0)[rc]{$\l_1$}}
\put(80.00,18.00){\makebox(0,0)[cc]{$\cdots$}}
\end{picture}
\caption{The extended tensor product graph for the product $V(\l_a)\otimes
V(\l_b)$ $(a\geq b)$ of two arbitrary fundamental representations of $C_n$.
The nodes correspond to representations whose highest weight is given by
the sum of the weight labeling the column and the weight labeling
the row. If $a+b>n$ then the graph extends to the right only up to
$\l_{n}$.\label{figtpg}}
\end{figure}
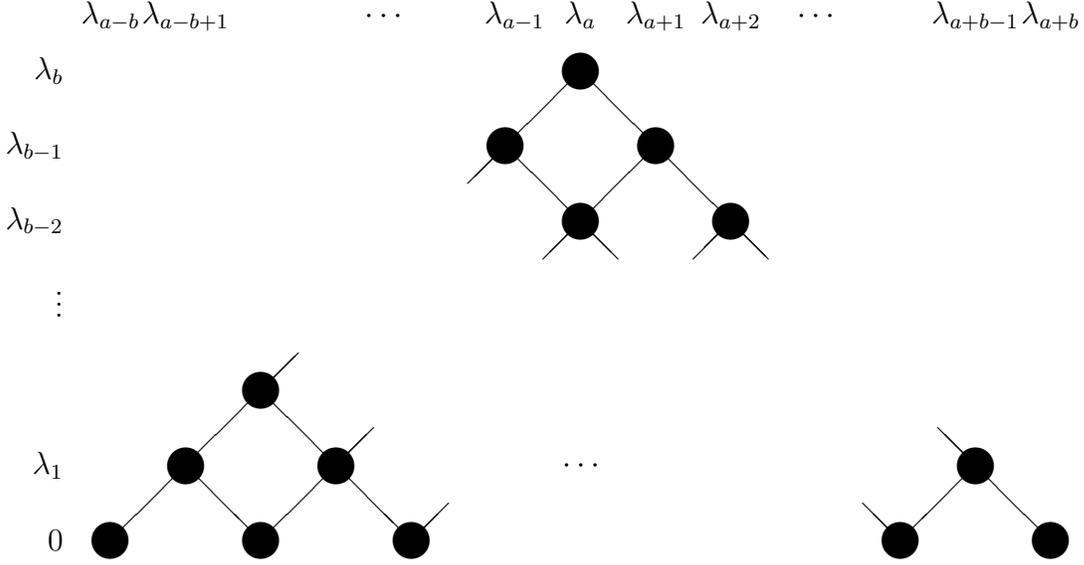

For $U_q(c_n^{(1)})$ the spectral decomposition of all the R-matrices
$\check{R}_{ab}$, with $\pi_a$ and $\pi_b$ being any two fundamental
representations, are known.
In the homogeneous gradation they were given in \ct{Hol94}, see
also \ct{Mac92,Del94b}.
Using the same notation as in \Eq{formr},\Eq{rhos}, they are
\[\label{cr}
\check{R}^{(h)}_{ab}(\th)=c_{ab}(\th)
\sum_{c=0}^{{\rm min}(b,n-a)}
\sum_{d=0}^{b-c}
\prod_{i=1}^c\langle a-b+2i\rangle \,
\prod_{j=1}^d\langle 2n+2-a-b+2j\rangle
\check{P}_{ab}^{(cd)},
\]
where by $(cd)$ we denote the irreducible $U_q(c_n)$ representation with
highest weight $\l_{a+c-d}+\l_{b-c-d}$. Without loss of generality we
have chosen $a\geq b$. This rather complicated formula is encoded in
the  ``extended tensor product graph'' displayed in figure \ref{figtpg}.
Each node in that graph corresponds to an irreducible $U_q(c_n)$
representation which appears in the tensor product of the two fundamental
representations $a$ and $b$. Thus they correspond to the intertwining
projectors $\check{P}$ in \Eq{cr}. The prefactor of a particular
$\check{P}$ is obtained as product of $\<l\>$ factors, one for
each link on a path from the corresponding node on the graph to the
top node. It turns out that the choice of path does not matter. The
integer $l$ in the $\<l\>$ factor corresponding to a particular link is
half the difference between
the values of the Casimirs of the connected nodes. The details of this
construction in the general case are described in \ct{Del94b}.

\subsection{The scalar prefactors}

Hollowood \ct{Hol94}, has found a prefactor $g_{ab}(\th)$ such that the matrix
\[
\tilde{S}_{ab}^{(h)}(\th)=g_{ab}(\th)\check{R}_{ab}^{(h)}(\th)
\]
satisfies the unitarity relation \Eq{unitarity} and the crossing
relations \Eq{crossing1},\Eq{crossing2} and, in addition,
has no poles on the physical strip.
For the details of this prefactor we refer the reader to Hollowood's
paper \ct{Hol94}.

$\tilde{S}_{ab}^{(h)}(\th)$ is in the homogeneous
gradation, we need to transform it to the gradation $(s)$ determined
by the Lorentz spins $\tilde{s}_i$ of the quantum group charges.
We have learned in section \ref{sectrmatrix} how to achieve this.
\[
\tilde{S}_{ab}^{(s)}(\th)=
\left(\pi_a(e^{\th H^{(s)}})\otimes 1\right)\tilde{S}_{ab}^{(h)}(\mu\th)
\left(1\otimes\pi_a(e^{-\th H^{(s)}})\right).
\]

Because of the absence of any
particle poles, this matrix can not yet be an S-matrix.
Rather it needs to be
multiplied by another prefactor $X_{ab}(\th)$
\[
S_{ab}(\th)=X_{ab}(\th)\tilde{S}_{ab}^{(s)}(z),
\]
$X_{ab}(\th)$ has to satisfy the S-matrix axioms by itself and has to
have particle poles at those values of $\th$ for which the R-matrix
$R_{ab}^{(s)}(\th)$
projects onto subrepresentations corresponding to another
fundamental representation $c$. We claim that such a prefactor is
given by
\[
X_{ab}(\th)=\prod_{p=1}^b\{a-b-1+2p\}_{\Htt}
\{\Htt-a+b+1-2p\}_{\Htt},
\]
where we use the notation
\[
\{a\}_{\Htt}=\frac{(a-1)_{\Htt} (a+1)_{\Htt}}{(a-1+B)_{\Htt}
(a+1-B)_{\Htt}},~~~
(a)_{\Htt}=\frac{\sinh\left(\frac{\th}{2}+\frac{i\pi}{2\Htt}a\right)}
{\sinh\left(\frac{\th}{2}-\frac{i\pi}{2\Htt}a\right)}.
\]
$\Htt=\htt-B$ and $B$ is a parameter which we will be related to
$q$ and $\mu$. For $c_n^{(1)}$ the \dkC number is $\htt=2n+2$.

These factors $X_{ab}$ are nothing else but the scalar S-matrices of
the fundamental particles of $d_{n+1}^{(2)}$ Toda theory which were
found in \ct{Del92b}. In that reference the pole structure
of these $X_{ab}$ for $0<B<2$ has
been investigated and all poles on the physical strip have been
shown to either be particle poles or to
arise from the Coleman-Thun mechanism. In particular some
simple poles are shifted away from their single-particle position by
higher order processes. The remaining true particle poles
were used to check consistency with the bootstrap principle.

Thus the only thing which remains to be checked is that at the
particle poles of $X_{ab}(\th)$,
$\tilde{S}_{ab}^{(s)}(\th)$ projects onto submodules.
If $a+b\leq n$, $X_{ab}(\th)$ has a particle pole at
\[
\th_{\rm pole}=i\pi\frac{a+b}{\Htt},
\]
We read off from \Eq{cr} that the residue of $\check{R}_{ab}^{(h)}(\th)$
at $\th=[\o](a+b)$ projects onto $V_{a+b}$. Correspondingly,
$\tilde{S}_{ab}^{(s)}$ projects onto $V_{a+b}$ at $\th=\frac{[\o]}
{\mu}(a+b)$. Thus if we set
\[
\Htt=i\pi\frac{\mu}{[\o]}
\]
then indeed the particle pole in $X_{ab}$ corresponds to the propagation of a
particle of type $a+b$. Using the constaint \Eq{cc} coming from crossing
symmetry we see that this implies that
$\Htt$ is related to the \dkC number $\htt$ by
\[
\Htt=\htt+\frac{2\pi i m}{[\o]},~~~~~\mbox{i.e., }
B=-\frac{2\pi i m}{[\o]}.
\]
{}From the location of the pole we can
calculate the quantum masses of the particles up to an overall
scale $M$ by using formula
\Eq{massc}. We find
\[\label{qm}
M_a=M\sin\left(\frac{a\pi}{\Htt}\right).
\]

The simple poles in $X_{ab}$ at $\th=i\pi\frac{\Htt-a-b}{\Htt}$ are due to
the propagation of the same particle type $a+b$ but in the crossed channel.
The simple poles in $X_{ab}$ with $a+b>n$ at $\th=i\pi\frac{a+b}{\Htt}$
are not particle poles. Rather the single particle poles are shifted
by higher order processes as been explained in \ct{Del94b}.

The reason why Hollowood in \ct{Hol94} was not able to find a
consistent S-matrix was that he was implicitly working with a
gradation which corresponds to $\Htt=\htt$ and at this particular point
the prefactors $X_{ab}$ which we have found reduce to $1$.

\subsection{The breathers\lab{sectbreathers}}

There is one more set of simple poles which we have not discussed yet and
which lie on the physical strip only if $B$ is negative.
In the $d_{n+1}^{(2)}$ Toda theory at real coupling constant, $B$
is positive and thus these poles have not appeared in \ct{Del94b}.
If $B$ is negative these simple poles lead to more states in the
spectrum of the
$\uqc$ symmetric theory and in analogy to affine Toda theory we will
call these bound states  ``breathers''.

$X_{aa}(\th)$ has a single pole at
\[
\th_{\rm pole}=i\pi\frac{\htt}{\Htt}.
\]
At this value of $\th$, $\tilde{S}_{ab}^{(s)}$ projects onto the trivial
one-dimensional representation $V_0$, as can be seen from \Eq{cr}.
Thus two solitons of type $a$ can create a breather singlet state of mass
\[
\left(M^{\rm breather}_a\right)^2
=2M_a^2\left(1+\cos(\pi\frac{\htt}{\Htt})\right).
\]
The pole in $X_{aa}(\th)$ at $\th=i\pi\frac{-B}{\Htt}$ is due to the
propagation of the same breather state in the crossed channel.
The S-matrices describing the scattering of these breathers could
be obtained from the $S_{ab}$ by applying the bootstrap.

More poles will occur for $|B|>1$, leading to further excited particle
states

\section{Conclusion}\lab{sectconclusion}

Quantum affine algebras have been shown to be a practical tool to
construct exact 1+1 dimensional relativistic S-matrices which satisfy
all the axioms of S-matrix theory. This has been explicitly demonstrated
by an example.

By a careful study of the consequences
of the Lorentz spins of the symmetry charges and the requirement of
crossing symmetry we have found the formula
\Eq{poleloc2} for the location of the particle poles which eventually
determine the quantum masses. It is pleasant to see a certain
``quantum'' dual Coxeter number to appear in this formula, mirroring
the way in which a ``quantum'' Coxeter number appeared in the pole
locations of the fundamental particles of affine Toda theory.

We expect that these S-matrices will find applications in several
1+1 dimensional quantum field theories, in particular as the soliton
S-matrices of quantum affine Toda theory. This will allow the further
study in these theories of such properties as the strong-coupling ---
weak-coupling duality, the breather --- particle duality and the
algebra --- dual algebra duality.

{\bf Acknowledgements:}

This paper is a product of my interest in the
problem of the soliton S-matrices for non-selfdual Toda theories.
I would like to thank the following people and others for discussions
about this problem and of quantum grops (in chronological order):
Marc Grisaru, Patrick Dorey, Tim Hollowood, Sergei Khoroshkin,
Valerij Tolstoy, Tony Bracken, Mark Gould, Yao-Zhong Zhang,
David Olive and the Swansea group, Mike Freeman, Ed Corrigan and the
Durham Group, Gerard Watts, Nial MacKay and Andrew Pressley.

\end{document}